\documentstyle[aps,prl,psfig,floats]{revtex}
\def\update{1 May 2000}
\def\version{[rev18 -- {\bf nucl-th/0005016]}}

\tighten

\def\fs#1{#1\!\!\!/}
\def\ja{{\hat{\jmath}}}
\def\sca{{\sc a}}
\def\scw{{\sc w}}
\def\scp{{\sc p}}
\def\sch{{\sc h}}

\begin{document}
\preprint{}
\draft

\wideabs{

\title{Pion photo- and electroproduction and the partially-conserved axial current}

\author{Helmut Haberzettl}

\address{Center for Nuclear Studies, Department of Physics, 
         The George Washington University, Washington, D.C. 20052}

\date{\update}

\maketitle

\begin{abstract}
The relevance of the axial current for pion production processes off the nucleon with real or virtual photons 
is revisited. Employing the hypothesis of a partially-conserved axial current (PCAC), it is shown that,
when all of the relevant contributions are taken into account, PCAC does not provide any additional
constraint for threshold production processes that goes beyond the Goldberger--Treiman relation.
In particular, it is shown that pion electroproduction processes at threshold cannot be used to extract any
information regarding the weak axial form factor. The relationships found in previous investigations
are seen to be an accident of the approximations usually made in this context.
\end{abstract}
\pacs{PACS numbers: 13.60.Le, 11.40.Ha, 14.20.Dh \hfill {\tiny \version}}
}

The hypothesis of a partially-conserved axial current (PCAC) \cite{weinberg2} has been employed in many investigations
for constraining scattering processes involving pions at threshold. One of its early successes was the relation 
by Goldberger and Treiman \cite{gt58} between the strength of the weak decay of the
nucleon $g_\sca$ and the strong-interaction $\pi NN$ coupling constant $g_{\pi NN}$, i.e.,
\begin{equation}
\frac{g_\sca}{f_\pi} \approx \frac{g_{\pi NN}}{m}\;,
\label{gtapprox}
\end{equation}
where $f_\pi$ is the weak decay constant of the pion and $m$ is the nucleon mass.
Experimentally this relation is found to be satisfied to better than 10\%.

Recently, PCAC relations were employed to extract the properties
of the nucleon's weak decay axial form factor $G_\sca$ from threshold pion electroproduction data
(see \cite{choi93,mainz99}, and references therein).
These extractions are based on the assumption that $G_\sca$ is related to the electromagnetic
structure of the Kroll--Ruderman contact term 
\cite{krollruderman54,ns62,adler65,debaenst70,er72,vz72,sk91}. 

I will  show here that the previous derivations of this relationship are based on an incomplete evaluation of
the relevant PCAC expressions and that, if all mechanisms are taken into account, the dependence
on $G_\sca$ vanishes. Thus, it will become obvious that the
identification of $G_\sca$ as the form factor entering the Kroll--Ruderman term is an accident
of the usually employed approximations.

To set the stage, the basic PCAC relations \cite{weinberg2} will be recapitulated first.
Excluding `second class' (i.e., tensor) currents, 
the general  form of the weak axial current is given by
\begin{eqnarray}
j^\mu_\sca &=&     - \overline{u}_f \gamma_5\Big[ \gamma^\mu G_\sca
                  +(p-p')^\mu G_{\scp} \Big]\frac{\tau}{2}  u_i \;.
\label{jax}
\end{eqnarray}
The PCAC hypothesis constrains this current by
\begin{eqnarray}
(p'-p)_\mu j^\mu_\sca &=& - \frac{ f_\pi \mu^2}{t-\mu^2}  \overline{u}_f \gamma_5 {G}_t u_i \tau \; ,
\label{pcac}
\end{eqnarray}
which provides a conserved current for vanishing pion mass $\mu$.
$G_t$ is the $\pi NN$ vertex function; other than the $\gamma_5$ which has been pulled out explicitly,
I make no assumptions about the internal structure of $G_t$.
Of course, within the present context, i.e., between on-shell spinors, 
$G_t$ is a function of $t=(p-p')^2$ only, where $p$ and $p'$ are the initial and final nucleon
momenta, respectively. $\tau$ is the vertex isospin operator. 
Here and throughout the present work, a Cartesian isospin basis is being employed, and 
the corresponding indices are suppressed; summation over these indices is implied when quantities carrying 
isospin indices are multiplied with each other.

The two form factors of the axial current are related via Eq.\ (\ref{pcac}), i.e., 
\begin{equation}
2m G_\sca +t G_\scp = -2 f_\pi \frac{\mu^2}{t-\mu^2} G_t\;.
\label{gtreleq}
\end{equation}
Evaluated at $t=0$, this provides the Goldberger--Treiman relation \cite{gt58},
\begin{equation}
 \frac{G_\sca(0)}{f_\pi} = \frac{G_t(0)}{m} \;.
\label{gt1}
\end{equation}
Equation\ (\ref{gtapprox}) assumes
that $G_\sca(0) \approx G_\sca(\mu^2) \equiv g_\sca$ and $G_t(0)\approx G_t(\mu^2) \equiv g_{\pi NN}$
since the pion mass is small.

While strictly speaking, Eq.\ (\ref{pcac}) is presumed to be valid 
only for $t$ values up to order $\mu^2$ \cite{weinberg2},
I will in the following take all of the preceding relations at face value,
assuming them to be valid at the operator level, and will consider
limits of small $t$, etc., only at the end.

Introducing an operator $\ja^\mu_\sca$ for the axial current, i.e., 
\begin{equation}
j^\mu_\sca= \overline{u}_f \ja^\mu_\sca u_i\;,
\label{jax1op}
\end{equation}
it can be split into weak and hadronic parts according to
\begin{equation}
\ja^\mu_\sca=\ja^\mu_{\sca,\scw}+\ja^\mu_{\sca,\sch}\;, \label{ja}
\end{equation}
where, having eliminated $G_\scp$ with the help of Eq.\ (\ref{gtreleq}),
\begin{mathletters}\label{jaxdef}
\begin{eqnarray}
\ja^\mu_{\sca,\scw} &=& -    \gamma_5\left[  \gamma^\mu
                   +(p'-p)^\mu\frac{2m}{t}  \right] {G}_\sca  \frac{\tau}{2}
            \;,\label{jax2}
\\
\ja^\mu_{\sca,\sch} &=&  - f_\pi (p'-p)^\mu \frac{\mu^2}{t} \frac{1}{t-\mu^2} 
                    \gamma_5 G_t \tau \;\label{jax3}
\end{eqnarray}
\end{mathletters}
separate the dependence on the weak and hadronic form factors, respectively. 
The divergence of the weak part,
\begin{eqnarray}
(p'-p)_\mu \ja^\mu_{\sca,\scw} &=&    \Big[\gamma_5(\fs{p}-m)+(\fs{p}'-m)\gamma_5\Big]
\,{G}_\sca \frac{\tau}{2} \;,
\label{divw0}
\end{eqnarray}
which vanishes between nucleon spinors, signifies the conserved part of the current and
\begin{eqnarray}
(p'-p)_\mu \ja^\mu_{\sca,\sch} &=&  -f_\pi \frac{\mu^2}{t-\mu^2} \gamma_5 G_t \tau 
\label{divh0}
\end{eqnarray}
provides the PCAC divergence of Eq.\ (\ref{pcac}).

\begin{figure}[t!]
\centerline{\psfig{file=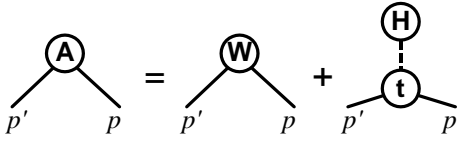,width=.7\columnwidth,clip=,silent=}}
\vspace{3mm}
\caption[text]{\label{figax} 
Splitting of the axial current $\ja^\mu_\sca$ into a conserved weak part 
$\ja^\mu_{\sca,\scw}$ and a pion-pole-dominated hadronic part $\ja^\mu_{\sca,\sch}$;
the latter produces the PCAC divergence of Eq.\ (\ref{pcac}). Here, and in all other 
diagrams, time proceeds from right to left.}
\end{figure}

Note that the two contributions $\ja^\mu_{\sca,\scw}$ and $\ja^\mu_{\sca,\sch}$ may be 
interpreted as resulting from the two diagrams of Fig.\ \ref{figax}. The hadronic current 
$\ja^\mu_{\sca,\sch}$, in particular, provides the straightforward interpretation of the 
pion-pole-dominated diagram of Fig.\ \ref{figax}: It describes the creation of the pion of 
mass $\mu$, with coupling operator $-f_\pi (p'-p)^\mu$ and associated normalized `form factor' 
$\mu^2/t$, and the subsequent propagation of the pion and its final absorption in the nucleon. 
In other words,
\begin{eqnarray}
\ja^\mu_{\pi} &=&   - f_\pi  \hat{q}^\mu  \frac{\mu^2}{\hat{q}^2}
\label{h1}
\end{eqnarray}
corresponds to the circle labeled {\sf H} in Fig.\ \ref{figax}, with $\hat{q}=p'-p$
being the pion's four-momentum flowing out of {\sf H}. 

The preceding operator-level description of the axial current, and its diagrammatic interpretation,
will provide precise meaning for the following of how the photon couples to the axial current.

\begin{figure}[b!]
\centerline{\psfig{file=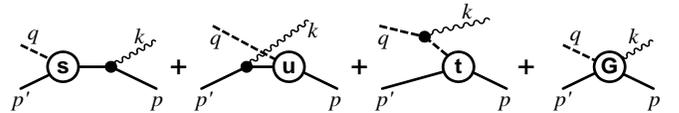,width=\columnwidth,clip=,silent=}}
\vspace{3mm}
\caption[text]{\label{figpi} 
Pion photoproduction for real or virtual photons. 
The last diagram marked {\sf G} depicts the interaction current 
$M^\nu_{\rm int}$; it subsumes the Kroll--Ruderman contact term, exchange-current contributions, and 
final-state interactions. The sum of all four diagrams is gauge-invariant
\cite{hh97g,hh00fsi}.}
\end{figure}

Turning now to the main issue of the present work, i.e., the production of pions off the nucleon with real or virtual photons,
the corresponding amplitude ${\cal M}$ is determined by the four diagrams in Fig.\ \ref{figpi}
\cite{hh97g,hh00fsi}, i.e.,
\begin{equation}
{\cal M} = \overline{u}_f \left( M^\nu_s + M^\nu_u + M^\nu_t + M^\nu_{\rm int} \right) u_i\, \varepsilon_\nu 
\;.\label{M}
\end{equation}
Adapting the PCAC hypothesis to pion photoproduction by employing minimal substitution,
Adler \cite{adler65} finds that ${\cal M}$ satisfies 
\begin{eqnarray}
\frac{f_\pi \mu^2}{q^2-\mu^2} {\cal M}
 &=& 
 q_\mu J^{\mu\nu}_{\sca,{\gamma}}\varepsilon_\nu
 -Q_\pi j^\nu_\sca \varepsilon_\nu \;,
\label{max}
\end{eqnarray}
where $J^{\mu\nu}_{\sca,{\gamma}}$
describes the coupling of the photon to the axial current and
$j^\nu_\sca$ is the nucleon matrix element (\ref{jax1op}) of the axial current.
$(Q_\pi)_{kl}=ei\varepsilon_{k3l}$ is the pion charge operator.
Note that only the nucleons are on-shell here, but the pion is off-shell.

This relation between the pion photoproduction amplitude ${\cal M}$ and the axial
current is presumed to be valid only in the limit of vanishing pion momentum $q$.
In the soft-pion limit $q\to 0$, 
following Ref.\  \cite{debaenst70}, the first term on the right-hand side here is often taken as zero
(see also \cite{er72,sk91}).
If true, this immediately provides
\[
\left.{\cal M}\right|_{q=0} = -  e_\pi \overline{u}_f \frac{\gamma_5\gamma^\nu}{2m}\widetilde{G}_\sca(k^2) u_i \varepsilon_\nu 
\;,
\]
where  $e_\pi=Q_\pi\tau$ effectively describes the charge of the outgoing pion.
The operator structure of this expression, $\gamma_5 \gamma^\nu$, is identical to the Kroll--Ruderman contact current
\cite{krollruderman54},
with a form factor
$\widetilde{G}_\sca(k^2) = G_t(0) G_\sca(k^2)/G_\sca(0)$
that derives its normalization from the $\pi NN$ form factor $G_t$ but its functional behavior
from the axial form factor $G_{\textsc{a}}$; $k$ here is the incoming photon momentum.
This is taken as evidence that the electromagnetic structure of the
Kroll--Ruderman term must be described by the axial form factor $\widetilde{G}_\sca$
and that $Q_\pi j^\nu_\sca \varepsilon_\nu$ may be used as the starting point for
extracting the threshold behavior of pion production processes
by considering expansions around $q=0$ \cite{debaenst70,er72,vz72,sk91}. 

In the following, I will show that these results follow from an incomplete treatment
of the prevailing dynamical situation and that none of these conclusions is warranted.
In doing so, it will become clear that Eq.\ (\ref{max})---which is based on the applicability
of the minimal substitution procedure to the present case---may need to be modified to
correctly describe the fact that the hadrons have internal structure.

To this end, I will consider the divergence
of the current $J^{\mu\nu}_{\sca,{\gamma}}\varepsilon_\nu$ of Eq.\ (\ref{max}).
Instead of evaluating this in the usual manner by the LSZ reduction scheme 
\cite{weinberg2,adler65,er72}, 
it is much more convenient to do this in terms of Feynman diagrams, consistent with
the operator approach adopted here for the axial currents.

\begin{figure}[t!]
\centerline{\psfig{file=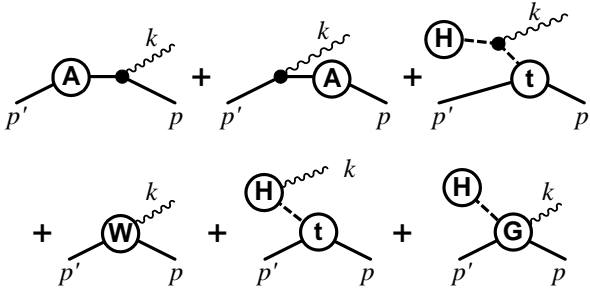,width=.9\columnwidth,clip=,silent=}}
\vspace{3mm}
\caption[text]{\label{figaxgam} Axial photoproduction current $\hat{J}^{\mu\nu}_{\sca,\gamma}$ of Eq.\ (\ref{dax}).
The diagrams are obtained from the axial current of Fig.\ \ref{figax} by inserting 
a photon line in all possible places.
In the top line, the circles denote the places where the axial current is evaluated:
{\sf A} corresponds to the full axial current of Fig.\ \ref{figax}, and {\sf H}
to its pion-pole-dominance part.
In the second line, the circles labeled {\sf W} and {\sf H} with attached photon
line correspond to contact operators $W^{\mu\nu}$ and $H^{\mu\nu}$, respectively.
The last diagram, where the photon is interacting within the $\pi NN$ vertex of the pion-pole
diagram of Fig.\ \ref{figax}, contains the interaction current $M^\nu_{\rm int}$
of pion photoproduction (cf.\ Fig.\ \ref{figpi}).}
\end{figure}

The current $J^{\mu\nu}_{\sca,{\gamma}}$ corresponds to inserting photon lines in 
all possible places in the axial-current diagrams of Fig.\ \ref{figax}.
The result is shown in Fig.\ \ref{figaxgam}; it can be verified either by direct inspection
of all relevant graphs or, in a more formal way, by using the gauge-derivative method 
of Ref.\ \cite{hh97g} (which is completely equivalent to the usual expansion of the relevant
Green's functions in terms of the electromagnetic field $A^\nu$ and summing up all
first-order contributions in $A^\nu$). In terms of operators, the resulting expression is
\begin{eqnarray}
\hat{J}^{\mu\nu}_{\sca,\gamma}
&=&
\ja^\mu_{\sca,f} \frac{1}{\fs{p}+\fs{k}-m} \Gamma^\nu_i 
+ \Gamma^\nu_f \frac{1}{\fs{p}'-\fs{k}-m} \ja^\mu_{\sca,i}
\nonumber\\
& &{ }
+\ja^\mu_{\pi} \frac{1}{q^2-\mu^2} \Gamma^\nu_\pi 
            \frac{1}{t-\mu^2}\gamma_5 G_t \tau + W^{\mu\nu} 
\nonumber\\
& &{ }
+  H^{\mu\nu}  \frac{1}{t-\mu^2}\gamma_5 G_t \tau
+\ja^\mu_{\pi} \frac{1}{q^2-\mu^2}  M^\nu_{\rm int}\;.
\label{dax}
\end{eqnarray}
The operators $W^{\mu\nu}$ and $H^{\mu\nu}$ describe the respective contact terms from 
the second line of Fig.\ \ref{figaxgam}. $H^{\mu\nu}$ is given by
\begin{eqnarray}
 H^{\mu\nu} &\equiv& -\left\{ \ja^\mu_\pi(p'-p) \right\}^\nu 
\nonumber\\
&=& -f_\pi \mu^2  \left[ g^{\mu\nu} -\frac{q^\mu (2q-k)^\nu}{q^2} \right] \frac{Q_\pi}{t}\;,
\label{hhdef}
\end{eqnarray}
where $-\{j^\mu_\pi\}^\nu$ is the gauge-derivative notation of Ref.\ \cite{hh97g}
which describes the coupling of the photon to $\ja^\mu_\pi$ of Eq.\ (\ref{h1}).
Note that this is identical to what one obtains from minimal substitution,
which is appropriate here
since $\ja^\mu_\pi$ does not contain any unknown functions.
By contrast, for $W^{\mu\nu}$, defined analogously as
\begin{equation}
W^{\mu\nu} \equiv -\left\{ \ja^\mu_{\sca,\scw}(p'-p) \right\}^\nu \;,
\label{wwdef}
\end{equation}
one cannot give a result in closed form in general since the internal structure of $G_\sca$
is unknown.
However, for the present discussion it suffices to know that $W^{\mu\nu}$ only depends on
$G_\sca$ since this is the only form factor contained in $\ja^\mu_{\sca,\scw}$.
[For a discussion of the structureless limit $G_\sca \to g_\sca$,
see remarks pertaining to Eq.\ (\ref{w0}).]

In Eq.\ (\ref{dax}), $\Gamma^\nu_\pi$ is the electromagnetic current for the pion
and the (gauge-invariant) current for the nucleon is
\begin{mathletters}
\begin{eqnarray}
\Gamma^\nu_{\sc n} &=& \gamma^\nu Q_{\sc n} + 
T^\nu_{\sc n}\;,
\\
T^\nu_{\sc n}&=&\left(\gamma^\nu k^2-k^\nu \fs{k}\right) \frac{F_1-1}{k^2} Q_{\sc n} 
+ i \frac{\sigma^{\nu\lambda} k_\lambda}{2m} \hat{\kappa}_{\sc n} F_2\;.
\label{jtrans}
\end{eqnarray}
\end{mathletters}
${\rm N}=i,f$ denotes the initial or the final nucleon;
$F_1$ and $F_2$ are the usual Dirac and Pauli form factors;
$Q_{\sc n}$ and $\hat{\kappa}_{\sc n}$ are the nucleon charge and anomalous magnetic moment
operators. Note that $\tau Q_i =e_i$ and  $ Q_f\tau =e_f$ provide effective 
(Cartesian-basis) charge operators for the nucleons in the present context and that one has
$e_i = e_f+e_\pi$,
describing charge conservation across the $\pi NN$ vertex.

Of particular importance in Eq.\ (\ref{dax})
is the interaction current $M^\mu_{\rm int}$ which originates from the photon attaching itself
within the $t$-channel $\pi NN$ vertex of the pion-pole-dominated diagram of Fig.\ \ref{figax}.
In lowest order (bare vertices), this corresponds to the usual gauge-invariance-preserving
Kroll--Ruderman term as obtained by
minimal substitution. In higher orders, with fully dressed vertices, this term contains a dressed
Kroll--Ruderman term, exchange currents, and 
all contributions from final-state interactions \cite{hh97g,hh00fsi}. 

In evaluating the divergence
\begin{equation}
(p'-p-k)_\mu \overline{u}_f \hat{J}^{\mu\nu}_{\sca,\gamma} u_i \varepsilon_\nu 
= - q_\mu J^{\mu\nu}_{\sca,\gamma} \varepsilon_\nu     \;,
\label{dmu}
\end{equation}
it is crucial to note that this involves divergences of the
axial current contributions $\ja^\mu_{\sca,f}$, $\ja^\mu_{\sca,i}$,
and $\ja^\mu_{\pi}$ according to Eqs.\ (\ref{divw0})-(\ref{h1})
which {\it do not vanish even when} $q\to0$.
The corresponding divergences of the first three and the last terms in Eq.\ (\ref{dax}),
in fact, produce the complete photoproduction amplitude ${\cal M}$,
plus electromagnetic contact terms arising from employing Eq.\ (\ref{divw0}).
Indeed, one now easily finds that
\begin{equation}
 q_\mu J^{\mu\nu}_{\sca,{\gamma}}\varepsilon_\nu
 -Q_\pi j^\nu_\sca \varepsilon_\nu 
= \frac{f_\pi \mu^2}{q^2-\mu^2} {\cal M} 
 +\overline{u}_f  {\cal W}^\nu u_i \varepsilon_\nu\;,
\label{maxw}
\end{equation}
where
\begin{equation}
{\cal W}^\nu =
q_\mu W^{\mu\nu} - Q_\pi \ja^\nu_{\sca,\scw}  -\frac{\gamma_5 \tau \Gamma^\nu_i + \Gamma^\nu_f \tau \gamma_5}{2} G_\sca(q^2) \;,
\label{wdep}
\end{equation}
with the last term containing the electromagnetic contact contributions.      
Equation (\ref{maxw}) is the desired final result and 
several remarks are in order now. 

To conform to Eq.\ (\ref{max}), ${\cal W}^\nu$ should vanish. However,
since Eq.\ (\ref{max}) was derived \cite{adler65} with the help of minimal substitution, 
this is only required in the structureless limit $G_\sca \to g_\sca$.
Evaluating $W^{\mu\nu}$ of Eq.\ (\ref{wwdef}) in this limit, one easily finds
\begin{equation}
W^{\mu\nu} \longrightarrow  -\gamma_5\left[g^{\mu\nu}-\frac{q^\mu(2q-k)^\nu}{q^2} \right] \frac{m}{t} g_\sca e_\pi\;,
\label{w0}
\end{equation}
which, employing also $F_1=1$ and $F_2=0$, indeed leads to ${\cal W}^\nu=0$ and thus verifies 
the validity of Adler's relation (\ref{max}) in this limit.
In general, however, for nucleons with electroweak structure, it is not obvious that this remains true.
One would need a microscopic description of the weak form factor $G_\sca$ to determine 
whether ${\cal W}^\nu$ still vanishes. 

The derivation of Eq.\ (\ref{maxw}) clearly shows that the entire $G_{\sc{a}}$ dependence of its left-hand side
is contained solely in ${\cal W}^\nu$ on the right-hand side.
In other words, the pion-production amplitude ${\cal M}$ itself does not depend on $G_\sca$.
Moreover, in view of the explicit expressions available for the axial photoproduction current 
$\hat{J}^{\mu\nu}_{\sca,\gamma}$, one finds that evaluating the limit $q \to 0$ in Eq.\ (\ref{maxw})
produces simply an identity, but does not provide a constraint that would permit
one to extract the threshold behavior of the pion-production amplitude independent of performing
that limit in  ${\cal M}$ itself.

For $q_\mu \hat{J}^{\mu\nu}_{\sca,\gamma}$,
in particular, one finds
\begin{eqnarray}
\left.q_\mu \hat{J}^{\mu\nu}_{\sca,\gamma}\right|_{q=0}&=&-f_\pi\left[M^\nu_{\text{int}}
            +\frac{\gamma_5\gamma^\nu}{2m}\frac{m}{f_\pi}G_\sca(k^2)e_\pi\right]
            +{\cal W}^\nu
\nonumber\\
&&{ }
+ f_\pi \frac{k^\nu}{k^2} \gamma_5\left[G_t(k^2)-\frac{m}{f_\pi}G_\sca(k^2)\right]e_\pi \;.
\end{eqnarray} 
The right-hand side here vanishes only in the extreme structureless limit,
where all electroweak and hadronic vertices are bare and the interaction current 
$M^\nu_{\text{int}}$ reduces to the Kroll--Ruderman contact term.
In general, however, it will be non-zero and, therefore,
the often used approximation \cite{debaenst70,er72,sk91} of assuming that $q_\mu J^{\mu\nu}_{\sca,{\gamma}}\varepsilon_\nu$ 
vanishes for $q\to 0$ is unjustified for physical hadrons with structure.  
[Technically, the incorrect limit is obtained if one consistently
reduces the entire axial current to its $\gamma_5 \gamma^\mu$ part
when evaluating Eq.\ (\ref{dmu}).]

It should be noted that Nambu {\it et al.} \cite{ns62} do not employ
this incorrect limit. Nevertheless, their results suffer from an indiscriminate interchange of the limits 
$q\to 0$ and $\mu \to 0$. Clearly, to obtain meaningful threshold results in the chiral
limit $\mu \to 0$, one must perform the limit $q\to 0$ first. 
In terms of Eq.\ (\ref{maxw}), the results of \cite{ns62} correspond to 
performing on the right-hand side first the $q$ limit and then the $\mu$ limit, 
and reversing this order on the left-hand side.
On the left-hand side, putting $\mu=0$ at the outset makes the hadronic part $\ja^\mu_{\sca,\sch}$ 
of the axial current vanish from the very beginning and destroys
PCAC [cf.\ Eq.\ (\ref{divh0})]. In other words, $q_\mu J^{\mu\nu}_{\sca,\gamma}$ is evaluated
by omitting the third and sixth diagrams from Fig.\ \ref{figaxgam}. These are the terms that would normally 
produce the $t$-channel pion-pole contribution and the interaction current.
The term $Q_\pi j^\nu_\sca$ on the left hand-side 
is then erroneously interpreted as supplying these two terms since it accidentally happens to have the same 
structure for $q=\mu=0$. In view of this incorrect treatment the resulting production current is not
gauge-invariant and, therefore, an additional {\it ad hoc} current was added in Ref.\ \cite{ns62} to repair 
this deficit. The present derivation shows that when performing the limits correctly, this should not
be necessary since the
photoproduction current that enters Eq.\ (\ref{maxw}) is gauge-invariant to start with.

It should be emphasized that the derivation of Eq.\ (\ref{maxw}) given here is completely model-independent. 
It hinges only on describing the weak axial current in terms of the operators defined in Eqs.\ (\ref{jaxdef}).
This corresponds to an effective Lagrangian description completely consistent
with the general form (\ref{jax}) of the axial current and with the PCAC hypothesis.
This consistency is necessary to avoid an incomplete or partial evaluation of all contributing
mechanisms which---since many of the terms have a deceptively similar structure---would lead to erroneous 
conclusions almost as a matter of course \cite{ns62,adler65,debaenst70,er72,vz72,sk91}.

The present considerations show that Eq.\ (\ref{maxw}) is devoid of any additional dynamical content that is 
not already part of the original pion-production amplitude. In fact, it simply provides an
alternative definition of the corresponding on-shell amplitude by the reduction formula
\begin{eqnarray}
{\cal M}= \lim_{q^2\to\mu^2} \frac{q^2-\mu^2}{f_\pi \mu^2}\overline{u}_f \Big[
q_\mu \hat{J}^{\mu\nu}_{\sca,\gamma}-Q_\pi \ja^\nu_\sca-{\cal W}^\nu
\Big]u_i \varepsilon_\nu
\;,
\end{eqnarray}
where on the right-hand side the dependence on $G_\sca$ cancels completely even before
the limit is taken. This, therefore, does not provide any constraint that goes beyond 
the original PCAC equation (\ref{gtreleq}) which led to the Goldberger--Treiman relation. 
In particular, there is no justification in modifying the Kroll--Ruderman term by multiplying 
it with the axial form factor when considering virtual photons with $k^2\ne 0$. 

Within PCAC, therefore, pion electroproduction data at threshold clearly cannot be interpreted
in terms of $G_\sca$, in contrast to what is commonly believed.
How this can be reconciled with the findings of chiral perturbation theory
\cite{bernard92} remains an open question at present.

This work was supported in part by Grant No.\ DE-FG02-95ER-40907 of 
the U.S. Department of Energy.

\end{document}